\documentclass[manuscript]{acmart}

\usepackage{amsmath}
\usepackage{soul}
\usepackage{natbib}
\usepackage{subcaption}
\usepackage{wrapfig}
\usepackage{url}
\usepackage{mathtools}
\AtBeginDocument{%
  }

\setcopyright{acmcopyright}
\copyrightyear{2023}
\acmYear{2023}
\acmDOI{XXXXXXX.XXXXXXX}

\acmJournal{JDS}
\acmVolume{37}
\acmNumber{4}
\acmArticle{111}
\acmMonth{8}

\begin{document}

\title[OutRank: Speeding up AutoML-based Model Search]{OutRank: Speeding up AutoML-based Model Search for Large Sparse Data sets with Cardinality-aware Feature Ranking}

\author{Bla\v{z} \v{S}krlj}
\email{bskrlj@outbrain.com}
\orcid{1234-5678-9012}
\affiliation{%
  \institution{Outbrain inc.}
  \streetaddress{New York}
  \city{New York}
  \country{US}
}

\author{Bla\v{z} Mramor}
\email{bmramor@outbrain.com}
\affiliation{
  \institution{Outbrain inc.}
  \streetaddress{New York}
  \city{New York}
  \country{US}
}

\acmArticleType{Research}

\begin{abstract}
  The design of modern recommender systems relies on understanding which parts of the feature space are relevant for solving a given recommendation task. However, real-world data sets in this domain are often characterized by their large size, sparsity, and noise, making it challenging to identify meaningful signals. Feature ranking represents an efficient branch of algorithms that can help address these challenges by identifying the most informative features and facilitating the automated search for more compact and better-performing models (AutoML). We introduce OutRank, a system for versatile feature ranking and data quality-related anomaly detection. OutRank was built with categorical data in mind, utilizing a variant of mutual information that is normalized with regard to the noise produced by features of the same cardinality. We further extend the similarity measure by incorporating information on feature similarity and combined relevance. The proposed approach's feasibility is demonstrated by speeding up the state-of-the-art AutoML system on a synthetic data set with no performance loss. Furthermore, we considered a real-life click-through-rate prediction data set where it outperformed strong baselines such as random forest-based approaches. The proposed approach enables exploration of up to 300\% larger feature spaces compared to AutoML-only approaches, enabling faster search for better models on off-the-shelf hardware.
   
\end{abstract}
\maketitle

\section{Introduction and problem overview}
\label{sec:overview}
The design of modern recommender systems relies on the integration of different context types into complex feature spaces that enable high-quality recommendations~\cite{batmaz2019review}. Real-world data sets in this domain are often characterized by their large size, sparsity, and noise, making it challenging to identify meaningful signals efficiently. Models that are part of real-life recommender systems, on the other hand, need to be lightweight and efficient due to the scale at which they are deployed~\cite{sheikhpour2023local}. For this purpose, contemporary machine learning model/design flows are utilized and consist of multiple steps, including feature (pre)selection, model search/optimization and refinement~\cite{shah2017recommender}. A commonly considered light-weight type of algorithms used in recommender systems are factorization machine-based models~\cite{rendle2010factorization}. Albeit versatile in production, these methods require computationally expensive procedures for finding optimal single or combined features that comprise the final model. Such compact and well-performing model configurations can be found by utilizing \emph{automated search systems}~\cite{he2021automl,zheng2023automl} (AutoML). These systems are capable of exploring both the space of algorithms and their configurations and have become ubiquitous in the design of performance factorization machines~\cite{selsaas2015affm}. The caveat associated with considering AutoML systems is their \textbf{computational complexity} - search for optimal feature configurations is time-consuming and often requires \emph{dedicated hardware} (or clusters of machines). This process can be, to some extent, complemented with a computationally more approachable branch of \emph{feature ranking} algorithms~\cite{venkatesh2019review,wang2013online,zhao2019maximum}. \textbf{Feature ranking} is a technique that can help address these challenges by identifying the most informative features and thus speeding up AutoML. These algorithms have found their use across multiple domains, spanning from biomedicine~\cite{liu2017analysis} to recommender systems~\cite{zhao2019maximum}. They enable fast estimation of \textbf{features' relevances} (and their interactions), and can be considered as a form of \emph{prior information} for an AutoML-based model search.

The remainder of this paper describes key properties of OutRank, a system for versatile feature ranking and data-quality-related anomaly detection. OutRank utilizes a variant of mutual information that is normalized with regards to the noise produced by features of the \emph{same cardinality}, which in our use-cases outperforms other strong baselines, such as random forest-based ranking. Furthermore, the presented approach can scale to data sets comprising hundreds of millions of instances using commodity hardware. Complemented by probabilistic cardinality estimation and coverage profiling, OutRank enables efficient data pre-processing, profiling of new features, and pruning of the input feature space to speed up AutoML-based model searches. Further, we present \textbf{3MR} (Minimum redundancy, maximum relevance, maximum relation) heuristic that enables computation of fast non-myopic estimates of feature importances -- its utility is demonstrated on synthetic and real-life data, where it enables substantial speedups of an existing AutoML system (TPOT~\cite{parmentier2019tpot}) and is shown to offer better rankings compared to strong baselines such as Random forest-based~\cite{breiman2001random} rankings on an internal data set related to conversion rate prediction.

\section{System overview}
\label{sec:methods}

\begin{figure}[htb!]
    \centering
     \Description[An overview of OutRank showing data ingestion and subsequent processing.]{}
    \includegraphics[width=.7\linewidth]{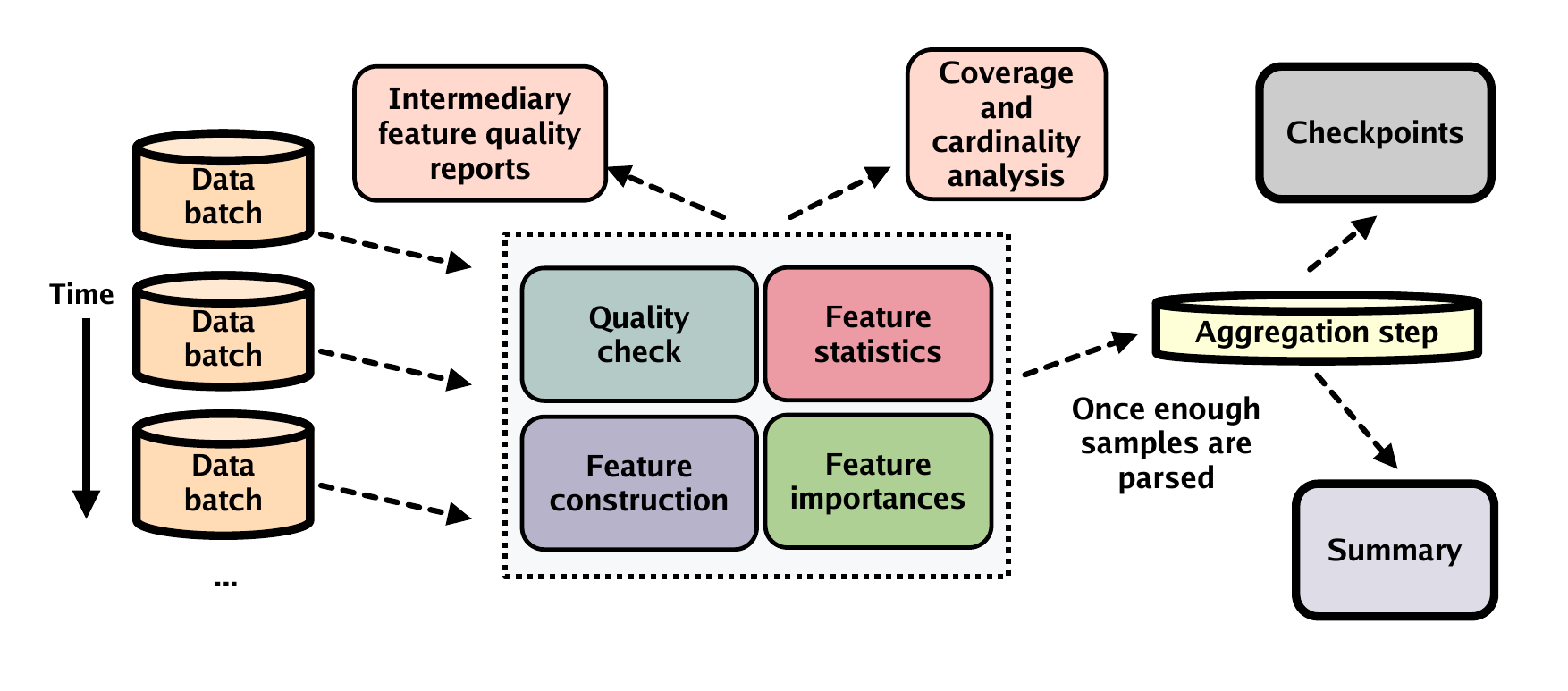}
    \vspace{-0.5cm}
    \caption{Overview of OutRank. Feature ranking and analysis are conducted for batches of data (the user parametrizes size) -- once enough data is observed by the system, the final aggregate is obtained from intermediary scores and used to produce final feature scores.}
    \label{fig:outrank-overview}
\vspace{-0.5cm}
\end{figure}

The remainder of this section is structured as follows. We first discuss the general context/overview of OutRank. Next, we discuss the impact of cardinality on ranking and how it was overcome. Finally, the section is followed by an overview of how feature combinations and their redundancies are considered in ranking, resulting in the 3MR algorithm.
We next discuss the system's architecture/overview. OutRank was built to facilitate the exploration of very large data sets (up to a billion instances in some cases) without any specialized hardware. The only resource-friendly approach to traversing and analyzing such data sets is by adopting batching -- the data set is streamed into the engine (OutRank). Once enough instances are temporarily stored, feature construction and ranking take place. This step is responsible for producing relevant transformations of data (on-the-fly), resulting in large space savings (compared to pre-computing all feature transformations of relevance), and conducting ranking. OutRank was built to scale to large data sets (tens of millions of listings) and support different ranking heuristics (discussed in the following sections). An overview of OutRank's architecture is shown in Figure~\ref{fig:outrank-overview}. 
This paper focuses on the novel contributions related to ranking features for sparse categorical data sets (a detailed discussion of OutRank's architecture is beyond the scope of this paper). However, the described overview should offer the reader the context in which such ranking is considered/used in practice.

We next discuss \textbf{overcoming the effect of high cardinality features}.
A common problem when dealing with data sets that are used for the construction of recommender systems is related to features' \textbf{cardinalities} -- the input space consists of mostly \emph{categorical features} that can be subject to varying amounts of unique (possible) values. Many unique values per feature are potentially problematic, as they decrease coverage per value (instances with a particular value) and can be harder to profile/assess when ranking for a downstream task such as modelling/transformation/further exploration. 
Mutual information (detailed overview in, e.g.,~\cite{kraskov2011erratum}) is an efficient similarity measure particularly suitable for comparing discrete (categorical) random variables. We next describe the variant of this measure we identified as fast and representative of features' similarities. The key motivation is based on the fact that the cardinality of a feature of interest plays an important role -- simultaneously, ranking features of different cardinalities can be sub-optimal and can result in biased final scores (we observed inflated scores of features that exhibit high cardinality). Based on the adopted notion of mutual information:
$$MI(U,V)=H(U) - H(U|V) = \mathbb{E}_{V} \big [ D_{\textrm{KL}} \big (p_{U|V=v} | p_U \big ) \big ],$$
where $U$ and $V$ represent two variables of interest and $D_{\textrm{KL}}$ represents Kullback-Leibler divergence, we propose the following modification.
To account for cardinality's impact, we observed that normalizing this score by the expected score obtained by aggregating MI scores across a collection of samples of a \textbf{random variable of same cardinality} has beneficial effects when differentiating between more and less important features. Hence, the mutual information-derived score we refer to as \textsc{CardMI} computed adheres to the following expression
\begin{align*}
\textsc{CardMI}(U, V) &=  \textsc{Aggregate}_{b \in B} \Big ( MI(U_b, V_b) - \mathbb{E}_{S_b \sim X_{|V_b|}} \big [ MI(U_b, S_b) \big ] \Big ), 
\end{align*}
\noindent where $B$ represents the set of all data batches (row-indexes, orange in Figure~\ref{fig:outrank-overview}) and $U,V$ represent two features under consideration. The $S_b$ represents a random sample of the same cardinality for a given batch of data (the number of such samples is a parameter). The aggregation function $\textsc{Aggregate}$ combines the intermediary scores into a final one and optionally truncates them to $[0,1]$ range. 
At the end of each run, resulting importance scores are scaled to 0-1 range (considering score vector $\textbf{x}$) $\textsc{Scale}(\textbf{x}_i) = \frac{\textbf{x}_i - \textrm{min}(\textbf{x})}{\textrm{max}(\textbf{x}) - \textrm{min}(\textbf{x})}.$
Intuitively, subtraction of the \emph{expected} noise is conducted to distil the signal at the batch level -- the generated noise differs from batch to batch (so does the actual data). Further, OutRank out-of-the-box implements a collection of \textbf{"sanity-check" features}; these features correspond to random noise of specific cardinality, constants and other types of noise for which expected ranking scores are known (and are either very low or high - if derived from the target variable directly). Comparison against controls of different cardinalities enables us to control the algorithm's performance with unit tests, offering fast insights into possible regression introduced during software updates. The main rationale for considering an MI-based score is its speed.
The computational complexity of \textsc{CardMI} is $\mathcal{O}(|F| \cdot b \cdot s \cdot |B|)$, where $s$ is the number of random samples used for normalization and $|B|$ batch size. Computing a single $MI$ estimate is fast for sparse data sets. Low computational complexity and \textbf{high parallelizability} within data batches (pairwise \textsc{CardMI} rankings) enabled the development of an extended algorithm that overcomes \textsc{CardMI}'s myopic nature -- lack of capability for accounting for interactions. This aspect is discussed next in more detail.

We next discuss \textbf{approximating higher order interactions}.
The crux of many recommender systems is their capability to accurately profile and detect relevant higher-order interactions. These \textbf{combinations} of two or more features can indicate sparse yet highly relevant events that govern the success of a given system. The adopted paradigm of factorization machines enables explicit modelling of such interactions; however, identifying them effectively is an ongoing research endeavour. Combining insights used by this branch of algorithms (\textbf{hashing trick}) with the efficiency of feature ranking, OutRank enables fast profiling of tens of thousands of interactions based on millions of instances. The idea underlying interaction profiling includes two insights that allow scaling. First, as OutRank traversed the data set in batches, profiling different interactions for different batches can substantially reduce time complexity ($\mathcal{O}(|F|^2)$). 
Each combined feature representing the interaction of multiple features is obtained by hashing existing value tuples (the data structure holding this information) into a single value, making the feature directly compatible with \textsc{CarMI} computation, i.e.,
    $\textsc{Comb}(F_1, F_2, \dots, F_n) = \textsc{Hash}(\textrm{struct}=(F_1, F_2, \dots, F_n)).$
\noindent Here, $\textsc{Hash}$ represents a fast hash function (xxHash in our case\footnote{\url{https://github.com/Cyan4973/xxHash}}). Albeit fast hashing enables the construction of combined features, ranking, in this case, takes substantially longer, and can be prohibitively complex. To overcome this issue, we introduce the notion of \emph{feature buffers} -- fixed-sized sets of features at the batch level that represent random samples of the space of all possible (generated) combinations. Feature buffer is \textbf{populated randomly per batch} (in an idempotent manner), and its fixed size guarantees consistent (predictable) performance. As different combinations are "sampled" when considering different batches, with enough data, reliable estimates of higher-order interaction scores are obtained (not enough samples imply unreliable scores).
This way, second and third-order (conjunctions) interactions are within reach for production-scale data sets. 
Apart from approximating the supervised effect of interactions (via hashing), OutRank can also compute lower-triangular similarity matrices that represent \textbf{redundancies} between pairs of features. This computational step is, similar to the supervised interaction one above, computationally expensive and is approximated in the same manner; only a subspace of possible combinations is considered per data batch - in the limit (=with enough data), enough samples of all combinations are obtained. In addition, the "buffer size", i.e. the number of combinations/features to be considered at most per batch, is parametrized - the lower the value, the more coarse-grained the approximation of similarities/interaction importance.

\subsection{3MR - Minimum Redundancy Maximum Relevance Maximum Relation extension}
\label{sec:modification}
Albeit able to account for cardinality-related issues better, the \textsc{CardMI} heuristic does not incorporate information about the feature's similarity to other features (it's myopic). To overcome this limitation, we introduce a computational step that recursively re-weights scores based on features' redundancies and relevance when they are present in interactions. The motivation for this comes from one of our major use cases for OutRank, speeding up the AutoML algorithms for model building. The idea/motivation is similar to mRMR~\cite{peng2005feature} and similar~\cite{zhao2019maximum} approaches but tailored to the use case of fast non-myopic feature ranking for categorical data. Further, more recent work on fast-mRMR~\cite{ramirez2017fast} also demonstrated the scalability of this branch of algorithms, which is aligned with our findings/design.
Re-weighting of features based on redundancies is the motivation behind the so-called MRMR heuristic~\cite{zhao2019maximum}. On the other hand, the reason for re-weighting features based on the interaction relevance comes from the fact that the recommendation algorithms which we most often use in AutoML (FM, FFM, DeepFM) favour features that, in interactions, generate strong signals. 

To describe this heuristic, let us denote by $\mathcal{S}_i \in \mathbb{R}$ with $i \in \{1,...,N\}$ the scores of the features $f_i$, $i \in \{1, ..., N\}$ computed by \textsc{CardMI} heuristic compared to the target. Furthermore, let $\mathcal{R}_{i,j} \in \mathbb{R}$ denote the \textsc{CardMI} scores where $R_{i,j}$ is the score of $f_i$ with respect to $f_j$. Finally, let $\mathcal{C}_{i,j} \in \mathbb{R}$ denote the \textsc{CardMI} scores of the interaction $f_i,f_j$ with respect to the target. Then the formula for the 3MR ranking can be given as a re-indexation (bijection) $F:\{1,...,N\} \to \{1,...,N\}$ such that $f_i$ is a higher ranked feature than $f_j$, if $i<j$. The heuristic can be computed iteratively by first considering $F(0) = \underset{i=1,\dots,N}{\mathrm{argmax}}(\mathcal{S}_i)$, followed by
\begin{align*} 
\eta = F(i) = \underset{j \notin F^{-1}\{1,\dots,i-1\}}{\mathrm{argmax}}\Big ( \overbrace{\mathcal{S}_{F(j)}}^\textrm{Feature's score} - \underbrace{\alpha \cdot \mathrm{SF}\left(\{\mathcal{R}_{j,k}\}_{k \in F^{-1}\{1,...,i-1\}}\right)}_\textrm{Redundancy information} + \underbrace{\beta \cdot  \mathrm{SF}\left(\{\mathcal{C}_{j,k}\}_{k \in F^{-1}\{1,...,i-1\}}\right)}_\textrm{Relational information}\Big ),
\end{align*}
where $\mathrm(SF)$ is a statistical function (e.g. mean, median, sum) and $\alpha, \beta > 0$ are hyper-parameters. Note that the set $F^{-1}\{1,...,i-1\}$ denotes all the features ranked above (better than) $i$ and in ($2$) this corresponds to all already computed features. Also, note that setting $\beta = 0$ brings us to the MRMR heuristic and setting $\alpha = 0$ makes the heuristic ignore redundancies and only consider relational information.

\section{Evaluation}
\label{sec:results}
\begin{figure}[t!]
     \centering
      \Description[Approximating AutoML generation with feature ranking is sensible and efficient (images showing alignment of that process with ranking results).]{}
     \begin{subfigure}[b]{0.46\textwidth}
         \centering
         \includegraphics[width=\textwidth, height=4cm]{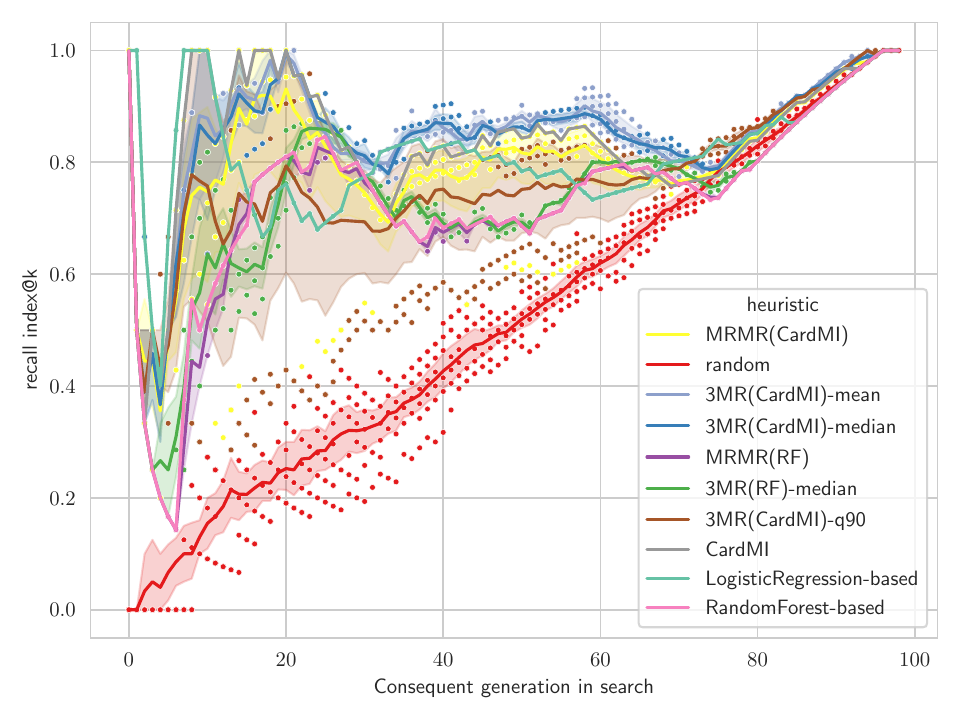}
         \caption{Approximating a single AutoML generation (single-feature factorization machines) with different ranking algorithms.}
     \end{subfigure}
     \hfill
     \begin{subfigure}[b]{0.46\textwidth}
         \centering
         \includegraphics[width=\textwidth, height=4cm]{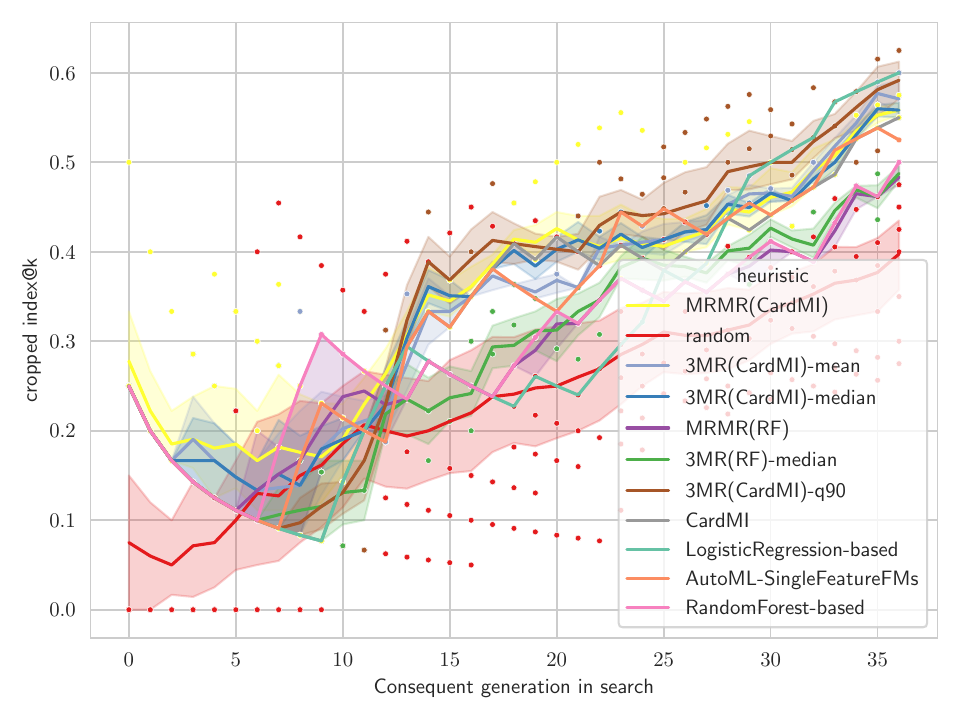}
         \caption{Approximating top features \emph{for each generation} of model search with different feature ranking algorithms.}
     \end{subfigure}
        \caption{Visualizations showing the potential of 3MR heuristic (computed as part of OutRank) for approximating computationally expensive in-house AutoML-based search. Subfigure (a) demonstrates that up to 20 features can be reliably retrieved via feature ranking. Subfigure (b) demonstrates that approximating the full trace of model search (multiple generations) is a hard problem, where proposed feature ranking heuristics perform better than AutoML if run only for a single generation.}
        \label{fig:small-data-bench}
\end{figure}

To evaluate the feasibility of the presented ideas, we conducted two sets of experiments designed to illustrate the methods' capabilities. First, we consider the common data mining scenario where a practitioner considers existing state-of-the-art AutoML tools for solving a given learning (classification in this case) task at hand -- we considered TPOT~\cite{parmentier2019tpot} system and explored \textsc{CardMI}'s capability to speed it up by pruning irrelevant part of the space before model search. The second part of the evaluation concerns using the presented work on a real-life production AutoML, where the amounts of data are greater, and the amount of time spent on feature search sufficient to identify suitable models can only be achieved in a distributed computing environment (>10 machines running model search).
We next discuss \textbf{performance on a synthetic data set - speeding up state-of-the-art AutoML}.
 The first experiment considers a well-known data set built to profile the quality of feature ranking algorithms -- Madelon~\cite{guyon2008feature}. The data set consists of 4{,}400 instances and 500 features; it is an artificial data set containing data points grouped in 32 clusters placed on the vertices of a five-dimensional hypercube and randomly labelled +1 or -1. 
 AutoML was re-run ten times (different random seeds) for each proportion of data (incrementally more features) to evaluate the performance/time relation of model search; each search was run for up to three generations with a population size of ten models. For this task, we used the \textsc{CardMI} computed in batches of 4196 instances -- this is one of the most straightforward off-the-shelf use cases applicable to many realistic scenarios. The results of more than two thousand AutoML runs are summarized in Figure~\ref{fig:small-data-bench}.
\begin{figure}[t!]
     \centering
      \Description[Two figures showing time-performance trade-offs.]{}
     \begin{subfigure}[b]{0.32\textwidth}
         \centering
         \includegraphics[width=\textwidth, height=4cm]{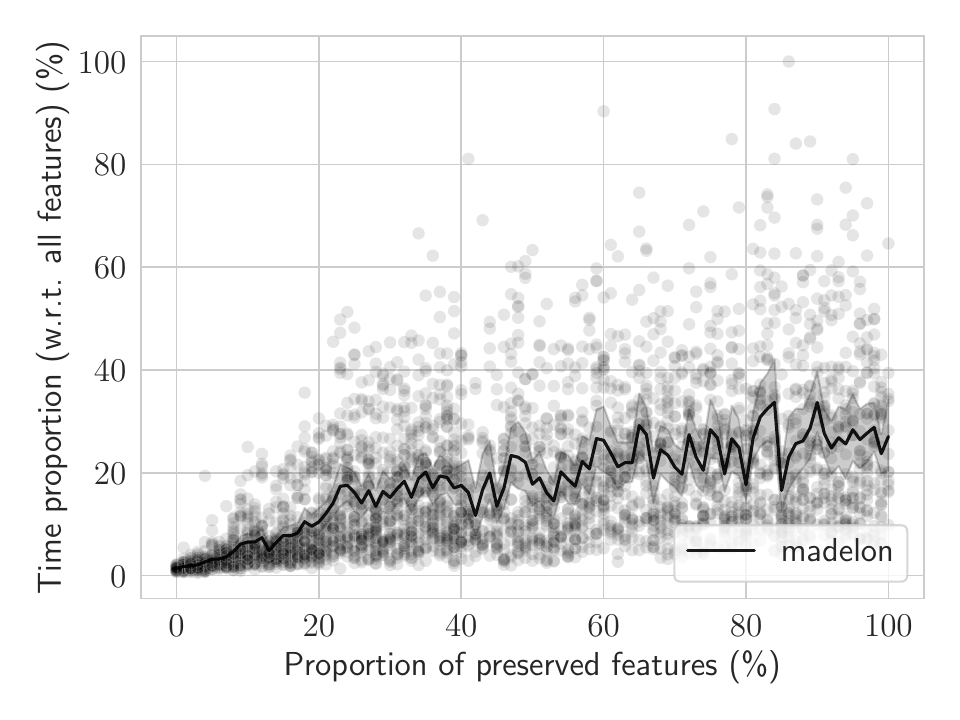}
         \caption{Model search time w.r.t. size of the (pre-pruned) feature space.}
     \end{subfigure}
     \hfill
     \begin{subfigure}[b]{0.32\textwidth}
         \centering
         \includegraphics[width=\textwidth, height=4cm]{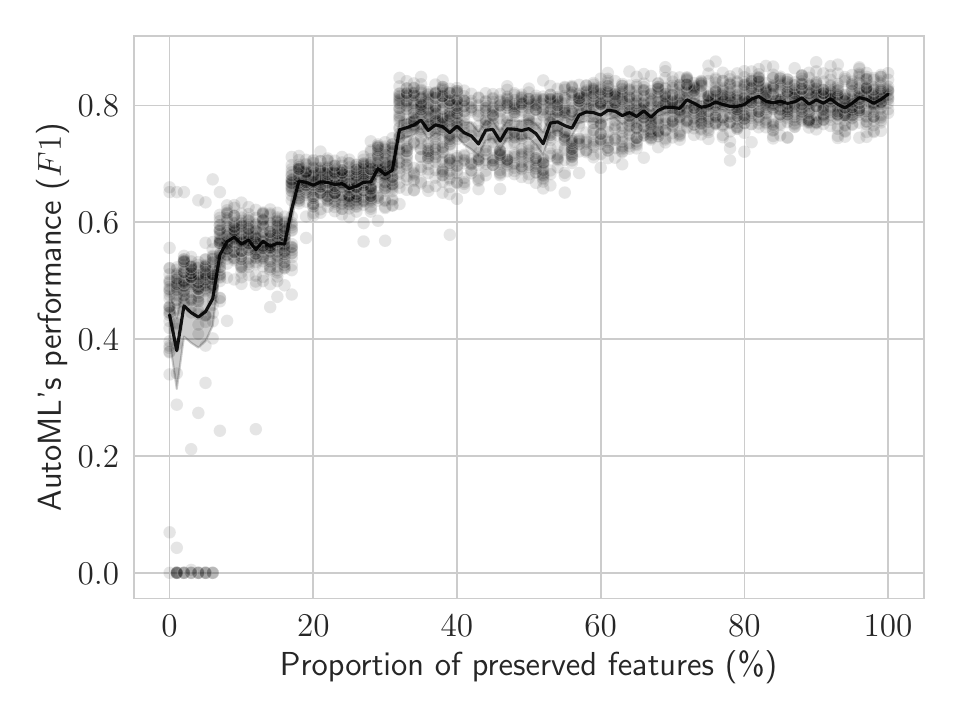}
         \caption{Relation between performances of best model and size of considered feature space.}
     \end{subfigure}
          \hfill
     \begin{subfigure}[b]{0.33\textwidth}
         \centering
         \includegraphics[width=\textwidth, height=4cm]{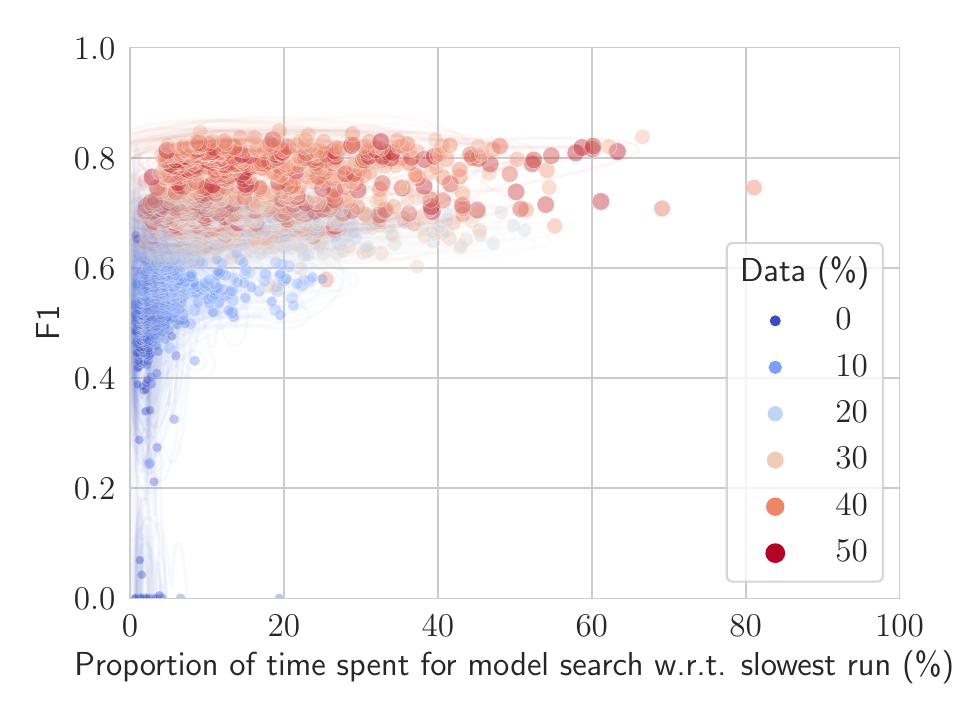}
         \caption{Relation between F1 performance and AutoML's time to achieve it.}
     \end{subfigure}
            \hfill
        \caption{Visualization of the impact of OutRank-based pruning on AutoML's performance. Visualizations show distributions of compute times and corresponding performances for various degrees of feature pruning (fewer features are considered right to left). Overall, more than 3{,}000 AutoML runs were conducted (one point=one run), and indicate that up to 30\% speedup can be obtained with minimal-to-no loss of performance (a,b,c). Further, considering all features from the get-go can have substantial memory overhead (requiring different hardware), offering another reason for fast feature pre-selection/denoising prior to more expensive AutoML-based model search.}
        \label{fig:toy}
\end{figure}
The benchmark (Figure~\ref{fig:toy}) illustrates the main motivation of this paper -- first, running more expensive AutoML was only possible after the initial feature "pre-selection". Second, on many data sets, up to 50\% time improvements were observed if considering smaller feature space (performance obtained by AutoML was within the margin of 0.5\% F1). The experiment demonstrates that efficient feature ranking can serve as a viable initial step when considering more expensive AutoML model search -- especially in higher-dimensional data sets, the performance benefits are substantial.
We proceed with the \textbf{performance on a real-life use case -- click-through rate (CTR) model search}.
Apart from data quality \emph{feature screening}, OutRank's feature ranking heuristics are valuable for speeding up our AutoML search. This can be done by dropping a proportion of irrelevant features and thus reducing the search space or by using top-ranked features as the \emph{seed model}, effectively \emph{skipping} multiple generations of expensive AutoML search. 
\begin{figure}[t!]
     \centering
      \Description[Performance of 3MR algorithm, it out-performs strong baselines on real-life data.]{}
     \begin{subfigure}[b]{0.46\textwidth}
         \centering
         \includegraphics[width=\textwidth, height=4cm]{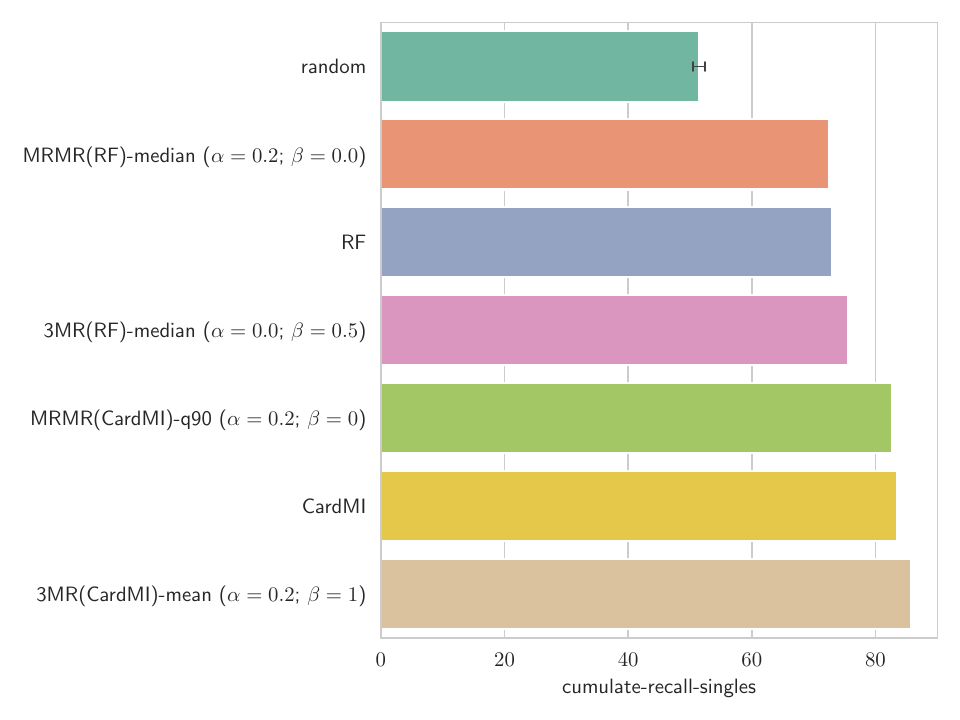}
         \caption{Performance of 3MR for the task of approximating single generation of search.}
     \end{subfigure}
     \hfill
     \begin{subfigure}[b]{0.46\textwidth}
         \centering
         \includegraphics[width=\textwidth, height=4cm]{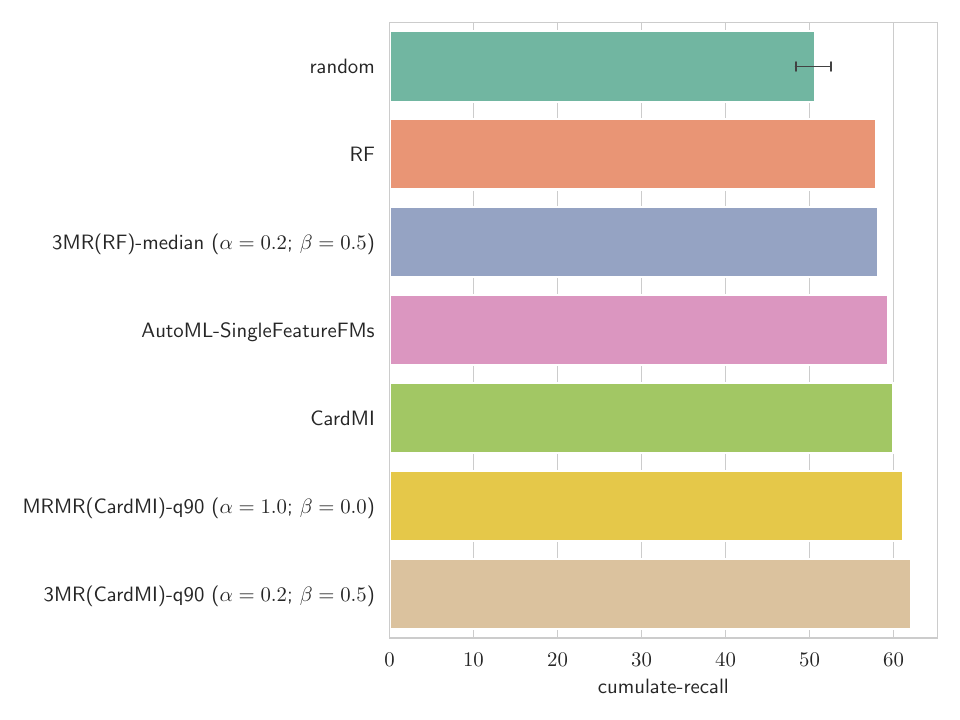}
         \caption{Performance of 3MR when approximating AutoML search trace (each generation).}
     \end{subfigure}
        \caption{Cumulative overview of the performance for two ranking tasks of interest. 3MR performs well for both tasks -- both redundancy and relation/interaction ($\alpha,\beta$)terms are non-zero, indicating that both types of information play a meaningful role during ranking.}
        \label{fig:barplots}
\end{figure}
To find optimal heuristics, we need a data set with a sufficient number of features and a benchmark feature ranking showing their importance for AutoML model-building procedure. Since such an open-source data set does not exist, we took 1.5 mio instances of subsampled production CTR training data with about 10\% positive labels and 100 features. Features were pre-selected to represent the heterogeneous nature of our data for the strength of their signal against the target, their cardinality, their correlations to other features and their number of missing values. An AutoML model search adding consecutively single features to a factorization machine gave us our main benchmark feature ranking (full AutoML run ranking). Another helpful feature ranking is the results of a single AutoML run, i.e. ranking based on the performance of a single-feature FM, which we currently use to speed up our search.

Data for \textbf{OutRank's feature ranking heuristics} was ten times subsampled AutoML data. The basic feature-ranking heuristics that we compared were logistic regression (LR), random forest (RF), randomly shuffled (R) and above mentioned CradMI. Furthermore, we compared MRMR and 3MR, which had the redundancies and relations computed with CardMI (due to higher computational complexity), and the relevances computed with RF or CradMI. The SF function in 3MR was mean, median or 90th percentile. The results can be seen in Figures  \ref{fig:barplots} and \ref{fig:small-data-bench}. 
For each $i = 1,...,100$ we compare ranking $G$ to a target ranking $F$ based on recall $R_i := \frac{\#\left(F^{-1}(1,...,i)\cap G^{-1}(1,...,i)\right)}{i},$ as in Fig.~\ref{fig:small-data-bench}. Furthermore, by taking the sum over $R := \sum_iR_i$, we get a general ranking for each heuristic compared to the benchmark, as in Fig.~\ref{fig:barplots}.
From Fig.~\ref{fig:barplots}a and b, we can see that against both benchmarks CardMI heuristic outperforms RF and that 3MR with the weight of the relation ($\beta$) larger than the redundancies weight ($\alpha$) further improves the ranking. For \emph{full AutoML run} benchmark, on the other hand, MRMR already has a better performance than CardMI, and 3MR is even better. For \emph{single AutoML run} benchmark MRMR is slightly worse, while 3MR with a large $\beta$ is better. Considering only redundancies gives a worse result in line with expectations since the single AutoML run benchmark is myopic by nature. Figure ~\ref{fig:small-data-bench} shows insights into performance of different heuristics on subsets of features.
The results confirm our intuition that adding redundancies and relations in the computation of feature ranking for AutoML improves the performance of the ranking heuristic. For our use-case, the proposed 3MR algorithm offers a good trade-off between speed/performance, even if compared against strong baselines.

\section{Conclusions and further work}
\label{sec:further-work}
This work explored whether simple hash-based encoding can offer added value in profiling interaction. Hash-based encoding is a simple and efficient technique that allows us to encode the interaction between two or more features into a single feature.
We demonstrated that \textsc{CardMI} heuristic, a variant of mutual information that incorporates features' cardinalities, offers pruned feature space that enables up to 30\% faster AutoML model search (TPOT) with no loss of performance and up to 50\% faster search with minimal performance loss ($\approx$ 0.5\% \textrm{F1}). Our results showed that even with simple hash-based encoding, we were able to achieve significant improvements in feature ranking compared to existing methods -- including strong baselines such as Random Forest-based ranking.
However, there are more complex schemes for encoding interactions that are also possible. For example, negations and conditionals instead of simple conjunctions can give rise to more complex features that are currently out of reach and might further boost the performance of feature profiling. The potential benefits of these more complex schemes for encoding interactions are clear, but they also come with added computational costs.
\textbf{The source of OutRank} will be made available upon concluded internal review required for its release\footnote{\url{https://github.com/outbrain/outrank}}.

\newpage
\bibliographystyle{ACM-Reference-Format}
\bibliography{sample-base.bib}

\end{document}